\documentclass[aps,pra,reprint,twocolumn,superscriptaddress,floatfix]{revtex4-2}

\usepackage{graphicx}
\usepackage{newtxtext,newtxmath}
\usepackage{mathtools}
\usepackage[colorlinks,linkcolor=blue,anchorcolor=blue,citecolor=blue,urlcolor=red]{hyperref}
\usepackage{orcidlink}
\usepackage{booktabs}
\usepackage{bm}
\usepackage{tikz}
\usetikzlibrary{quantikz2}
\usetikzlibrary{shapes.geometric}

\usetikzlibrary{decorations.pathreplacing}
\usetikzlibrary{shapes.geometric}

\newcommand{\casql}{Laboratory of Quantum Information, University of Science and Technology of China, Hefei, Anhui, 230026, China}
\newcommand{\casex}{Anhui Province Key Laboratory of Quantum Network, University of Science and Technology of China, Hefei 230026, China}
\newcommand{\aihf}{Institute of Artificial Intelligence, Hefei Comprehensive National Science Center, Hefei, Anhui, 230088, China}
\newcommand{\origin}{Origin Quantum Computing Technology (Hefei) Co., Ltd., Hefei, Anhui, 230026, China}

\begin{document}

\title{
Depth-reduced quantum simulation of Lindbladian dynamics via adaptive variational quantum trajectory compression
}

\author{Huan-Yu Liu\,\orcidlink{0000-0002-6158-9627}}
\email{liuhuanyu@ustc.edu.cn}
\affiliation{\aihf}

\author{Cheng Xue\,\orcidlink{0000-0003-2207-9998}}
\affiliation{\aihf}

\author{Yun-Jie Wang\,\orcidlink{0009-0000-5347-5286}}
\affiliation{\casql}
\affiliation{\casex}

\author{Xi-Ning Zhuang\,\orcidlink{0000-0001-5118-5066}}
\affiliation{\origin}

\author{Chao Wang}
\affiliation{\origin}

\author{Yu-Chun Wu\,\orcidlink{0000-0002-8997-3030}}
\email{wuyuchun@ustc.edu.cn}
\affiliation{\casql}
\affiliation{\casex}
\affiliation{\aihf}

\author{Zhao-Yun Chen\,\orcidlink{0000-0002-5181-160X}}
\email{chenzhaoyun@iai.ustc.edu.cn}
\affiliation{\aihf}

\author{Guo-Ping Guo\,\orcidlink{0000-0002-2179-9507}}
\affiliation{\casql}
\affiliation{\casex}
\affiliation{\aihf}
\affiliation{\origin}

\date{\today}

\begin{abstract}
Quantum simulation of open quantum systems in the noisy intermediate-scale quantum (NISQ) era is hindered by the non-unitary nature of dissipative dynamics and the limited quantum resources available on near-term quantum processors, namely restricted qubit counts and noise-limited circuit depths. In this work, we propose an efficient quantum algorithm for simulating Lindbladian dynamics on NISQ devices. For open quantum systems with Pauli dissipations, we first derive a compact and stable mixed-unitary adjoint channel that approximates the target dissipative dynamics and enables ancilla-free implementation through trajectory sampling. To further reduce the circuit depth required for implementing the sampled trajectories, we introduce an adaptive variational quantum trajectory compression framework. In this framework, a depth-adaptive parameterized quantum circuit is trained to approximate repeated Hamiltonian simulation operators, which are then used to replace repeated unitary segments appearing in the sampled trajectories. Importantly, the training procedure can also be performed without auxiliary qubits. Numerical simulations of the dissipative quantum $XY$ model and the interacting disordered $XXZ$ chain demonstrate the accuracy and reduced execution gate counts of the proposed algorithm. Our results provide a practical route toward ancilla-free and depth-reduced simulation of open quantum systems on near-term quantum hardware.
\end{abstract}

\maketitle
\section{Introduction}

Quantum computing has developed rapidly over the past few decades. Quantum algorithms, which offer the potential for speedups over their classical counterparts, have been proposed for a broad range of applications, including quantum chemistry~\cite{qccrmp, Cao2019qc, Bauer2020qc}, finance~\cite{Herman2023qcf, 9222275, liu2025qci}, machine learning~\cite{qmlnature, qmlfrature, qmlshadow, qmlhernal}, and the simulation of many-body quantum systems~\cite{qsimnature, qsimprx, qsimrmp}. In parallel, quantum processors based on platforms such as superconducting qubits~\cite{Wu2021advantageprl, superreview} and trapped ions~\cite{trappedion2022, trappedion2020, trappedion2026} have achieved continuous improvements in both scale and quality. Recent demonstrations of quantum error correction operating below the threshold~\cite{qec1, qec2} further herald the advent of fault-tolerant quantum computing. Nevertheless, the practical implementation of fault-tolerant quantum algorithms remains highly challenging. It is therefore essential to develop resource-efficient near-term quantum algorithms~\cite{neartermalg1, nisqalg} that can exploit the capabilities of currently available quantum hardware.

Although present-day quantum devices remain in the noisy intermediate-scale quantum (NISQ)~\cite{nisq} regime, processors with hundreds or even thousands of physical qubits are already available, making relatively large problem instances accessible in principle. In practice, however, their utility is primarily limited by quantum noise, which degrades operational fidelity and restricts the achievable circuit depth. This limitation is particularly severe for the quantum simulation of many-body dynamics, where time evolution is typically implemented through repeated Trotterized operations. From the perspective of near-term quantum computing, reducing algorithmic circuit depth is therefore a fundamental and urgent task.

A common strategy for reducing quantum resource requirements in the NISQ era is to incorporate classical computation, with variational quantum algorithms (VQAs)~\cite{vqareview} serving as a representative paradigm. In VQAs, parameterized quantum circuits (PQCs) are employed to prepare ansatz states and evaluate loss functions or gradients, while a classical optimizer updates the circuit parameters. Although it remains unclear whether VQAs can provide practical quantum advantage~\cite{liu2023can, vqslimit}, partly due to unfavorable trainability~\cite{trainnphard, bpnc} and expressibility~\cite{eacch1} issues, their reliance on shallow PQCs makes them particularly appealing for near-term quantum hardware.

In the context of circuit-depth reduction, a natural approach is variational quantum compiling (VQC)~\cite{vqc1, vqc2, vff1, hst1, ncvqc, vqcprr}, which approximates a target unitary operation using a shallow PQC. Based on this idea, variational quantum simulation (VQS) methods for closed-system dynamics have been widely proposed and demonstrated~\cite{vqsclosed1, vqsclosed2}. For example, Ref.~\cite{vff1} introduced a variational fast-forwarding algorithm based on VQC, where the time-evolution operator is approximately diagonalized by a trainable unitary so that long-time dynamics can be generated efficiently by modifying only the diagonal component. Ref.~\cite{vqsnc} proposed a stepwise strategy for encoding time evolution with a PQC, demonstrating ground-state search and the observation of phase transitions on superconducting quantum processors.

While isolated quantum systems follow unitary time evolution and constitute a central application domain of quantum computing, realistic quantum systems inevitably interact with their surrounding environments. The resulting open quantum systems~\cite{oqs1, oqs2} exhibit a variety of rich physical phenomena~\cite{Christopoulos2023ed, Purkayastha2016oed, Rakovszky2024steady, Jung1999pt, Nakanishi2022ptpt}. However, their quantum simulation is fundamentally more challenging because the underlying dynamics are generally non-unitary.

Quantum simulation of open quantum systems has been widely investigated~\cite{simulateoqsreview1, simulateoqsreview2}, and existing methods can be broadly classified according to how the non-unitary dynamics are implemented. One direct strategy is to embed the non-unitary evolution into unitary dynamics on an enlarged Hilbert space. Dilation-based approaches~\cite{Richard2019Efficient, Wang2011Quantum, Gaikwad2022Simulating, Ding2024Simulating, Han2021Experimental, Observation2021observation, jin2024sch, Shen2025Observation} construct a suitable dilation, implement the corresponding global unitary evolution, and then extract the desired dynamical information. However, these methods often face scalability challenges in both constructing the dilation and realizing the enlarged evolution, especially on NISQ devices. In addition, several works have shown that hardware noise can be exploited as a resource for realizing certain non-unitary dynamics~\cite{Bacon2001Universal, noiseass} and observing the ground-state properties~\cite{nafqa}. Nevertheless, such noise-assisted methods usually suffer from limited flexibility and substantial sampling overhead.

Another class of approaches decomposes open-system dynamics into an ensemble of quantum trajectories, each of which can be simulated independently and then combined through classical post-processing to recover the target dynamics. However, when individual trajectories involve non-unitary operations, auxiliary qubits are still required to implement their dilations~\cite{Quantum2021lcu}. The linear combination of Hamiltonian simulation approach~\cite{lchs, cbmd} provides an alternative route by decomposing non-unitary operations into weighted sums of unitary evolutions. As discussed in Ref.~\cite{vqlchs}, however, implementing the resulting summations through linear combination of unitaries (LCU) techniques~\cite{lcu}, or estimating the corresponding cross terms through measurements, is generally not well suited to current NISQ devices.

To avoid auxiliary qubits and the resulting deep circuits, several recent works have focused on simplified or structured models. Ref.~\cite{Peetz2024convex} proposed simulating mixed-unitary channels through sampling and averaging. Ref.~\cite{Borras2025quantum} considered the case of unitary dissipation and derived a mixed-unitary representation of the dynamics, enabling ancilla-free simulation. In our previous work~\cite{Liu2025simulationofopen}, we derived a compact adjoint channel that approximates the target dynamics, where the desired evolution can be obtained by simulating the adjoint channel followed by classical post-processing.

VQS methods for dissipative dynamics have also been explored. Ref.~\cite{vqsprl} introduced a general VQS framework in which the parameters encoding the quantum state are optimized according to variational principles. We extended the stepwise procedure of Ref.~\cite{vqsnc} to non-Hermitian dynamics~\cite{vqlchs}. Ref.~\cite{robustvqs} combined the general VQS framework with trajectory unraveling to achieve higher-order robust simulation of the Lindblad master equation. However, the evaluation of loss functions or matrix elements in these methods still requires auxiliary qubits or circuits with growing depth, posing a challenge for NISQ devices.

In this work, we investigate ancilla-free quantum simulation with reduced trajectory circuit depth of open-system dynamics with Pauli dissipations on NISQ devices. Building on our previous work~\cite{Liu2025simulationofopen}, we derive a compact and more stable adjoint channel that approximates dissipative dynamics without requiring the additional post-processing step. This channel can be efficiently simulated through trajectory sampling without using auxiliary qubits. To further reduce the circuit depth of the sampled trajectories, we introduce an adaptive variational quantum trajectory compression framework, in which a depth-adaptive PQC is trained to approximate repeated Hamiltonian simulation operations appearing in the compact adjoint channel. The trained PQC can then be used to replace repeated unitary segments appearing in the sampled trajectories, thereby reducing the circuit depth while preserving the sampled-channel structure. Numerical simulations on typical dissipative models demonstrate the correctness of the proposed method and its capability for circuit-depth reduction. Our approach highlights the potential of jointly exploiting problem-specific dissipative structures and variational compression techniques to bridge the gap between open-system quantum simulation algorithms and their implementation on NISQ devices.

\section{Adjoint Channel for Lindbladian Dynamics}

\begin{figure*}[ht]
    \centering
    \includegraphics[width=\textwidth]{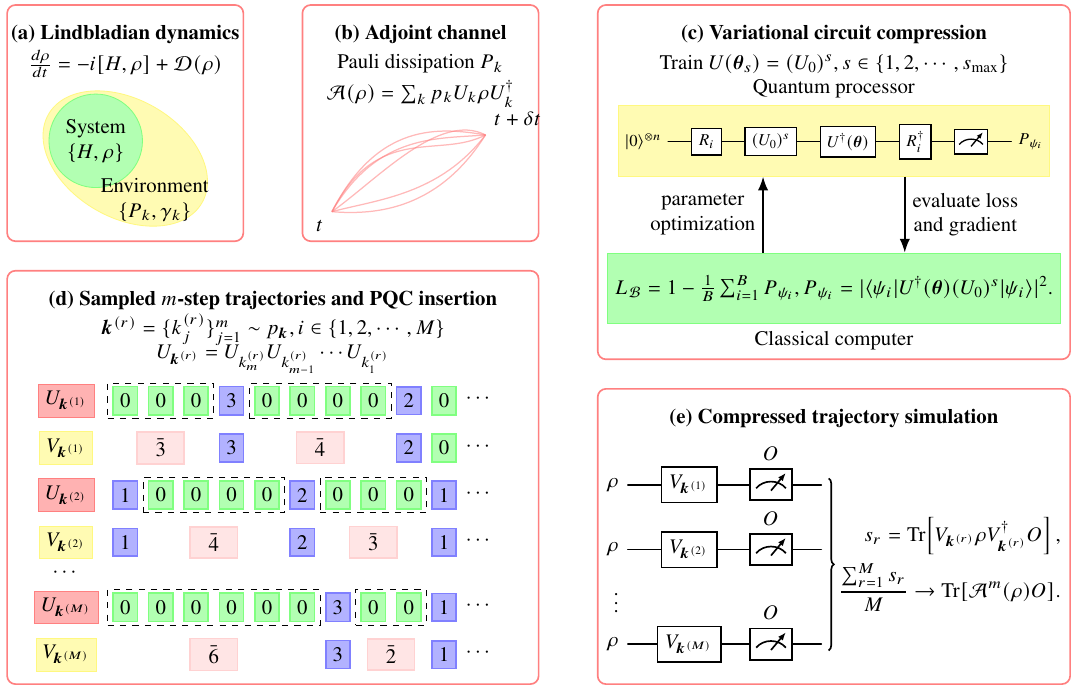}
    \caption{
    \textbf{Overview of the proposed simulation algorithm.}
    (a) Open quantum systems described by the Lindblad master equation, where the system of interest interacts with its surrounding environment.
    (b) For Pauli dissipations, the dissipative dynamics can be approximated by a compact mixed-unitary adjoint channel, enabling ancilla-free implementation through trajectory sampling.
    (c) Variational circuit compression framework. For each $s\in\{1,2,\cdots,s_{\text{max}}\}$, a depth-adaptive PQC $U(\bm{\theta}_s)$ is trained to approximate the repeated Hamiltonian evolution $(U_0)^s$. We evaluate the loss function on finite batches of random product states $\mathcal{B}$, each of which can be prepared via a series of single-qubit quantum gates $R_i=\otimes_{j=1}^n R_{ij}$, using a quantum processor, while a classical optimizer updates the parameters.
    (d) Monte Carlo sampling of $m$-step trajectories and PQC insertion. Each trajectory $U_{\bm{k}}$ is sampled according to the probability distribution $p_{\bm{k}}$, as shown by the red boxes. The green and blue boxes represent the circuit blocks corresponding to $U_0$ and $U_{k>0}$, respectively. Since $p_0$ is typically larger than $p_{k>0}$, consecutive $U_0$ blocks appear frequently in sampled trajectories. After the compression training in (c), each consecutive $U_0$ segment can be replaced by the optimized PQC $U(\bm{\theta}_s)$, indicated by the pink boxes labeled by $\bar{s}$, yielding the depth-compressed trajectory $V_{\bm{k}}$.
    (e) Simulation of compressed quantum trajectories. Each sampled and compressed trajectory is implemented independently, followed by the measurement of the observable $O$. Averaging the measurement outcomes over all sampled trajectories gives an estimate of the time-evolved expectation value $\operatorname{Tr}[\mathcal{A}^m(\rho)O]$.
    }
    \label{fig:overview}
\end{figure*}

In this paper, we focus on the dynamics of open quantum systems described by the Lindblad master equation~\cite{lindblad1, lindblad2}, where the system of interest interacts with its surrounding environment, as illustrated in Fig.~\ref{fig:overview}(a):
\begin{equation}\label{eq:lindblad}
    \frac{d\rho}{dt} = -i[H,\rho] + \mathcal{D}(\rho),
\end{equation}
with
\begin{equation}
    \mathcal{D}(\rho) = \sum_{k=1}^D \gamma_k \left( P_k \rho P_k^\dagger - \frac{1}{2}\{P_k^\dagger P_k,\rho\} \right),
\end{equation}
where $H$ and $\rho$ denote the Hamiltonian and density matrix of the system, respectively, while $\gamma_k$ and $P_k$ are the dissipation rates and the corresponding dissipation operators. To enable ancilla-free quantum simulation, we focus on Pauli dissipations with $P_k \in \{I,X,Y,Z\}^{\otimes n}$. Although this setting is specialized, Pauli dissipations have been widely studied~\cite{Peetz2024convex, Borras2025quantum, Liu2025simulationofopen, noiseass, Chen2018simulate} and arise in various relevant scenarios~\cite{Cai2023quantumem, van2023probabilistic, Brien2023Purification, Youngseok2023Scalable, Campbell2017roadsftqc, Babu2023errorcorrection, Sivak2023realtime, Miao2023overcoming}.

In our recent work~\cite{Liu2025simulationofopen}, we showed that such dynamics can be simulated by approximating the evolution with an adjoint channel, which is mixed-unitary and can therefore be implemented on a quantum processor without auxiliary qubits. The target dynamics are then reconstructed through classical post-processing. However, hardware noise and finite-sampling errors may be amplified during the post-processing stage, which can degrade the accuracy and stability of the method. To address this issue, we derive a more stable adjoint channel representation that eliminates the need for the additional classical post-processing used in the previous formulation.

For a time step $\delta t$, the quantum channel approximating Eq.~\eqref{eq:lindblad} with an error scaling as $\mathcal{O}(\delta t^2)$ can be written as
\begin{equation}
    \rho_{t+\delta t} = \sum_{k=0}^D M_k \rho_t M_k^\dagger,
\end{equation}
where the Kraus operators are given by
\begin{equation}
    M_0 = I - iH\delta t + K\delta t, \qquad
    M_{k>0} = \sqrt{\gamma_k \delta t}\, P_k .
\end{equation}
Denoting $\Gamma = \sum_{k=1}^D \gamma_k$, we have $K = -\frac{1}{2} \sum_{k=1}^D \gamma_k P_k^\dagger P_k = -\frac{1}{2}\Gamma I$, which ensures the normalization condition $ \sum_{k=0}^D M_k^\dagger M_k = I+\mathcal{O}(\delta t^2)$.

To obtain a mixed-unitary channel, each Kraus operator should be approximated by a unitary operator up to a scalar factor. For $k>0$, this condition is naturally satisfied because $P_k$ are unitary Pauli strings. It remains to treat $M_0$. In contrast to our previous work~\cite{Liu2025simulationofopen}, where the term $K\delta t$ was neglected, and the resulting deviation had to be corrected by classical post-processing, here we show that $M_0$ can be directly approximated by a rescaled unitary operator with an error scaling as $\mathcal{O}(\delta t^2)$:
\begin{equation}
\begin{aligned}
    M_0 &= I - iH\delta t - \frac{1}{2}\Gamma \delta t I \\
        &= \left(1-\frac{1}{2}\Gamma \delta t\right) \left( I - iH \frac{\delta t}{1-\frac{1}{2}\Gamma \delta t} \right) \\
        &= \left(1-\frac{1}{2}\Gamma \delta t\right) e^{-iH\frac{2\delta t}{2-\Gamma \delta t}} + \mathcal{O}(\delta t^2).
\end{aligned}
\end{equation}

Defining the rescaling factor $Z = \left(1-\frac{1}{2}\Gamma \delta t\right)^2$ and the rescaled time $\Delta t = \frac{2\delta t}{2-\Gamma \delta t}$, and performing a normalization, we obtain the adjoint channel, as illustrated in Fig.~\ref{fig:overview}(b),
\begin{equation}
    \mathcal{A}(\rho) = \sum_{k=0}^D p_k U_k \rho U_k^\dagger ,
\end{equation}
where
\begin{equation}\label{eq:adjoint_channel}
\begin{aligned}
    p_0 &= \frac{Z}{Z+\Gamma \delta t}, & p_{k>0} &= \frac{\gamma_k \delta t}{Z+\Gamma \delta t}, \\
    U_0 &= e^{-iH\Delta t}, & U_{k>0} &= P_k.
\end{aligned}
\end{equation}
This channel provides a mixed-unitary approximation to the target dissipative dynamics and can be simulated without auxiliary qubits.

For a multi-step simulation up to total time $t=m\delta t$, directly recording the time-evolved density matrix, for example through quantum state tomography, would be exponentially expensive. We therefore have to apply the channel $\mathcal{A}$ repeatedly to the initial state. The resulting $m$-step composite channel can be written as
\begin{equation}\label{eq:adjoint_m}
    \mathcal{A}^m(\rho) = \sum_{\bm{k}} p_{\bm{k}} U_{\bm{k}} \rho U_{\bm{k}}^\dagger ,
\end{equation}
where the multi-index is denoted by $\bm{k}=(k_1,k_2,\ldots,k_m)$, and
\begin{equation}
    p_{\bm{k}} = \prod_{i=1}^m p_{k_i}, \qquad
    U_{\bm{k}} = U_{k_m} U_{k_{m-1}} \cdots U_{k_1}.
\end{equation}

The total number of Kraus operators in this representation scales as $(D+1)^m$, making a direct summation over all trajectories computationally intractable for large $m$. Nevertheless, the action of $\mathcal{A}^m$ can be efficiently approximated by the Monte Carlo sampling procedure illustrated in Figs.~\ref{fig:overview}(d) and~\ref{fig:overview}(e). In each of the $M$ samples, a trajectory index $\bm{k}^{(r)} = ( k_1^{(r)}, k_2^{(r)}, \ldots, k_m^{(r)})$ is generated according to the probability distribution $p_{\bm{k}^{(r)}}$. The corresponding composite unitary $U_{\bm{k}^{(r)}}$ is then applied to the initial state $\rho$, followed by the measurement of an observable $O$. The time-evolved expectation value is estimated by averaging over $M$ independently sampled trajectories:
\begin{equation}\label{eq:sampleapprox}
    \frac{1}{M} \sum_{r=1}^M \operatorname{Tr} \left[ U_{\bm{k}^{(r)}} \rho U_{\bm{k}^{(r)}}^\dagger O \right] \to \operatorname{Tr} \left[ \mathcal{A}^m(\rho) O \right].
\end{equation}
According to the central limit theorem, the statistical sampling error scales as $\mathcal{O}(1/\sqrt{M})$, allowing the expectation value under the $m$-step channel to be estimated without explicitly enumerating the exponentially many Kraus trajectories. In the channel $\mathcal{A}$, the unitary $U_0$ corresponds to Hamiltonian evolution, whereas the operators $U_k$ with $k>0$ are Pauli strings. Consequently, the circuit depth of each sampled trajectory scales at most linearly with the total simulation time.

\section{Adaptive Variational Quantum Trajectory Compression}

For weak dissipation, namely small $\gamma_k$, and a sufficiently small time step $\delta t$, the no-jump probability typically dominates, i.e., $p_0 \gg p_k$ for $k>0$. Consequently, in a sampled trajectory indexed by $\bm{k}$, the Hamiltonian-evolution operator $U_0$ appears much more frequently than the jump operators $U_k$ with $k>0$. Although the adjoint-channel sampling strategy avoids the explicit summation over exponentially many Kraus trajectories, each sampled trajectory may still contain a large number of repeated applications of $U_0$ in long-time simulations. This leads to deep quantum circuits and poses a significant challenge for current NISQ devices. In practice, Hamiltonian simulation experiments on superconducting quantum processors are typically limited to about $20$--$30$ Trotter steps~\cite{ising127,hubbard72}, beyond which the simulation results can be severely degraded by hardware noise.

To mitigate this circuit-depth bottleneck, we propose an adaptive variational quantum trajectory compression strategy within the adjoint-channel sampling framework, as shown in Fig.~\ref{fig:overview}(c). The key idea is to train shallow PQCs to approximate repeated Hamiltonian-evolution blocks and then insert the trained circuits into sampled trajectories. Specifically, for each $s\in\{1,2,\cdots,s_{\text{max}}\}$, we train a PQC $U(\bm{\theta}_s)$ to approximate the $s$-step Hamiltonian evolution,
\begin{equation}\label{eq:pqccompress}
    U(\bm{\theta}_s) \approx (U_0)^s = \left(e^{-iH\Delta t}\right)^s .
\end{equation}

Since a global phase does not affect expectation values of the form $\operatorname{Tr}[U\rho U^\dagger O]$, the approximation between two unitary operators can be quantified by the phase-insensitive Hilbert-Schmidt fidelity (HSF),
\begin{equation}
    F_{\text{HS}}(U,V) =\left| \frac{1}{d} \operatorname{Tr}[U^\dagger V] \right|^2 ,
\end{equation}
where $d=2^n$ is the Hilbert-space dimension. The corresponding loss function is
\begin{equation}\label{eq:losshs}
    L_{\text{HS}}(\bm{\theta}) = 1 - \sqrt{F_{\text{HS}}  \left( U(\bm{\theta}), (U_0)^s  \right)}.
\end{equation}

The training objective can be evaluated using the Hilbert-Schmidt test (HST), as in \cite{hst1,vff1}, which requires two $n$-qubit registers; see Appendix~\ref{sec:hst} for details. Although HST directly evaluates the desired fidelity, the need for an additional register makes it less favorable for near-term implementations.

While the HSF reflects the agreement between the ansatz and target within the entire Hilbert space, and thereby requires an additional quantum register, some works considered evaluating this agreement by a finite batch of sampled states. Recent studies~\cite{ncvqc,vqcprr} have provided theoretical motivation and numerical evidence for training quantum dynamics using random product states. Following these works, we can also perform the training on products of independent single-qubit states $|\psi_i\rangle = \otimes_{j=1}^n |\psi_{ij}\rangle$ with each $|\psi_{ij}\rangle$ Haar-distributed and can be efficiently prepared via a single-qubit $U(3)$ gate as $|\psi_{ij}\rangle = R_{ij}|0\rangle$ \cite{ncvqc}. For a batch $\mathcal{B}=\{|\psi_i\rangle\}_{i=1}^B$, the agreement on each state $|\psi_i\rangle$ is
\begin{equation}
    P_{\psi_i}   = \left| \langle \psi_i|  U(\bm{\theta})^\dagger (U_0)^s  |\psi_i\rangle  \right|^2 ,
\end{equation}
which can be measured by the simple quantum circuit shown in Fig.~\ref{fig:overview}(c). Then the batch fidelity becomes
\begin{equation}
    F_{\mathcal{B}}  \left(  U(\bm{\theta}),(U_0)^s \right) = \frac{1}{B} \sum_{i=1}^B  P_{\psi_i}.
\end{equation}
The corresponding loss function is
\begin{equation}\label{eq:lossb}
    L_{\mathcal{B}}(\bm{\theta}) = 1  - F_{\mathcal{B}}  \left(  U(\bm{\theta}),  (U_0)^s \right).
\end{equation}
Since random product states contain coherent superpositions of computational-basis states, the sampled objective is more sensitive to relative phases while preserving an ancilla-free implementation, providing a hardware-friendly surrogate for training the compressed Hamiltonian-evolution blocks.

More importantly, the aim of circuit compression is to obtain a shallower approximation of the target unitary. It is not restricted to specific compilation strategies. Therefore, existing VQC strategies \cite{vqc1, vqc2, vff1, hst1, ncvqc, vqcprr}, non-variational compression algorithms \cite{kokcu2023}, as well as other circuit optimization strategies can be applied. Moreover, the structure of the adjoint channel further allows the compressed circuit library to be reused across trajectories.

Beyond the loss function choice, constructing the target circuit for $(U_0)^s$ is also important. Here we consider two training strategies. The first is a direct training strategy, where the PQC $U(\bm{\theta}_s)$ is trained to approximate the $s$-step Hamiltonian evolution $(U_0)^s$ directly. On quantum hardware, we would implement the target by repeatedly applying a circuit approximating $U_0$. This strategy provides a direct approximation to the desired unitary and avoids error accumulation between different values of $s$. However, the depth of the target circuit used for loss evaluation grows linearly with $s$, which can become costly when large compression lengths are required.

To avoid using a deep target circuit during training, we further introduce an iterative training strategy. We first train a shallow circuit $U(\bm{\theta}_1)$ to approximate a single no-jump step $ U(\bm{\theta}_1) \approx U_0$. After $U(\bm{\theta}_{s-1})$ has been trained as an approximation to $(U_0)^{s-1}$, the next circuit $U(\bm{\theta}_s)$ is trained to approximate
\begin{equation}
    (U_0)^s  =  (U_0)^{s-1} U_0  \approx  U(\bm{\theta}_{s-1}) U_0 .
\end{equation}
In this iterative procedure, the target circuit at each training stage consists of the previously compressed circuit followed by only one additional Hamiltonian-evolution step. Therefore, the circuit depth could be reduced compared with the direct training setting. The price one pays is that approximation errors may accumulate through the iterative construction. Thus, the choice between direct and iterative training, or a combination thereof, depends on the target simulation time, the available circuit depth, and the desired accuracy.

After the PQCs are trained, they can be used to compress the sampled trajectories, as illustrated in Fig.~\ref{fig:overview}(d). For a sampled trajectory indexed by $\bm{k}$, the corresponding composite unitary $U_{\bm{k}}$ can be decomposed into alternating products of jump blocks and consecutive no-jump blocks,
\begin{equation}
    U_{\bm{k}} = W_{l+1}(U_0)^{r_l}W_l \cdots  W_2(U_0)^{r_1}W_1 ,
\end{equation}
where $r_a>0$ denotes the length of the $a$-th consecutive no-jump block, and each $W_a$ is a product of jump operators $U_k$ with $k>0$. The boundary blocks $W_1$ and $W_{l+1}$ are allowed to be the identity, depending on whether the trajectory begins or ends with a no-jump segment.

The compression is then applied blockwise. If $r_a\leq s_{\text{max}}$, the block $(U_0)^{r_a}$ is replaced by the trained PQC $U(\bm{\theta}_{r_a})$. If $r_a>s_{\text{max}}$, we decompose the length as $r_a = c_a s_{\text{max}} + r_a'$ with $0\leq r_a' < s_{\text{max}}$, and replace the long no-jump block with
\begin{equation}
    (U_0)^{r_a}
    \approx \left[ U(\bm{\theta}_{s_{\text{max}}}) \right]^{c_a}
    \begin{cases}
        U(\bm{\theta}_{r_a'}), & r_a'>0,\\
        I, & r_a'=0.
    \end{cases}
\end{equation}
For notational convenience, we denote the compressed approximation to $(U_0)^{r_a}$ by $\bar{U}_{r_a}$, which is represented by $\bar{r}_a$ in the pink boxes in Fig.~\ref{fig:overview}(d). The depth-compressed trajectory is then written as
\begin{equation}
    V_{\bm{k}} = W_{l+1}\bar{U}_{r_l}W_l  \cdots  W_2\bar{U}_{r_1}W_1 .
\end{equation}
The compressed trajectories can subsequently be simulated and averaged according to Eq.~\eqref{eq:sampleapprox}, as also shown in Fig.~\ref{fig:overview}(e).

For the compression to be beneficial, the depth of $\bar{U}_{r_a}$ should be smaller than that of the original block $(U_0)^{r_a}$. To this end, we employ a depth-adaptive ansatz. Starting from a shallow circuit, the ansatz depth is increased only when the optimization fails to reach the desired fidelity. This adaptive strategy avoids unnecessarily deep circuits and strikes a practical balance among trainability, approximation accuracy, and circuit depth.

On the other hand, the training of VQAs may still be affected by optimization difficulties, such as local minima~\cite{trainnphard} and barren plateaus~\cite{bpnc}, which can reduce training efficiency and increase the required measurement overhead. To alleviate this issue, we can adopt a transfer-learning-inspired parameter initialization strategy, as discussed in our previous work~\cite{liu2023mitigating}. Specifically, when the optimization proceeds from time $t$ to $t+\delta t$, or from compression length $s$ to $s+1$, the optimized parameters obtained in the previous step are used as the initial parameters for the next optimization. Such a warm-start strategy exploits the continuity between neighboring target unitaries and can improve the stability and efficiency of the variational training, while mitigating the barren plateaus issue.

The choice of the maximum compression length $s_{\text{max}}$ also plays an important role in the overall efficiency of the algorithm. For small $s_{\text{max}}$, the target unitary $(U_0)^s$ is relatively shallow, and the corresponding loss function is easier to evaluate and optimize. However, only short consecutive no-jump segments can be compressed, leading to a limited reduction in circuit depth. In contrast, a larger $s_{\text{max}}$ allows longer Hamiltonian-evolution segments to be compressed into PQCs, but the direct evaluation of the training objective becomes more costly, and the iterative strategy may suffer from accumulated approximation errors. Therefore, $s_{\text{max}}$ should be chosen by balancing the available quantum resources and the desired simulation accuracy. 

The optimization cost should also be considered. For the quantum circuit used to evaluate the batch-sampled fidelity in Fig. \ref{fig:overview}(c), the state preparation and uncomputation operations $R_i$ and $R_i^\dagger$ consist only of single-qubit gates. The main contributions to the circuit depth come from the target $(U_0)^s$ and the ansatz $U(\bm{\theta})$. In direct training, the target circuit depth scales with $s$, which is bounded by the chosen maximum compression length $s_{\mathrm{max}}$, rather than by the total number of simulation steps $T/\delta t$. Together with the finite batch size and depth-adaptive ansatz, this makes the training procedure feasible.

However, it should be emphasized that reducing the execution depth of sampled trajectories does not necessarily imply a reduction in total computational resources when the training overhead is included. The primary objective of our framework is to obtain shallower circuits that can be executed coherently on depth-constrained NISQ devices, rather than to guarantee lower overall computational costs. This is consistent with existing VQC approaches, where additional optimization effort is employed to obtain shallower implementations of target unitary operations. Our framework extends this idea by constructing a reusable circuit library that can be applied across different stochastic trajectories, thereby avoiding the need to compile each trajectory independently.

Overall, the proposed adaptive variational compression framework converts frequently occurring repeated Hamiltonian-evolution blocks into shallow trainable circuits, thereby reducing the depth of sampled trajectories while preserving the ancilla-free structure of the adjoint-channel simulation.

\section{Numerical simulations}

To demonstrate the feasibility of the proposed framework, we perform numerical simulations of the dissipative $XY$ model and the interacting disordered $XXZ$ chain. The simulations in this section use noiseless numerical evaluations without measurement shot noise or hardware noise.

\subsection{Simulation details}\label{sec:simdetail}

We consider two training objectives: the full-space HSF and the sampled fidelity over a finite batch of random product states. The finite-batch training results are presented in this section, and those with the full-space objective are shown in Appendix \ref{sec:suppfull}. In our numerical simulations, $U_0$ is evaluated exactly by diagonalizing the Hamiltonian. Both objectives are investigated using direct and iterative training. Direct training uses $(U_0)^s$ as the target, while iterative training employs the previously optimized PQC and another $U_0$.

In contrast, the uncompressed trajectory circuits used for gate-count comparisons are implemented using Trotter decompositions, as detailed in Appendix~\ref{sec:simcirc}. On quantum hardware, evaluating the training loss also requires a circuit implementation of the target, which may introduce additional approximation errors.

For full-space training, we employ the L-BFGS-B optimizer, with the step size determined adaptively by line search. The convergence tolerances are $\mathrm{ftol}=10^{-12}$ and $\mathrm{gtol}=10^{-8}$, with at most 20 line-search steps per iteration. Each optimization attempt, corresponding to a fixed depth and initialization, is limited to 100 iterations and may terminate earlier when a convergence criterion is satisfied. We retain the parameters yielding the highest HSF among all objective evaluations. After completing the prescribed initialization attempts at a given depth, we accept the depth if the best HSF reaches $0.99$ for direct training or $0.999$ for iterative training. Otherwise, we proceed to the next candidate depth within the prescribed search range.

For finite-batch training, we use $B=32$ and the Adam optimizer with a fixed learning rate $\eta=0.02$, $(\beta_1,\beta_2)=(0.9,0.999)$, and $\epsilon=10^{-8}$. At each update, a fresh batch of 32 random product states is generated independently. Each optimization attempt consists of 100 updates. A separate, fixed set of 256 random product states is used to evaluate the parameters after every five updates. We retain the parameters with the highest average fidelity on this validation set. The depth-acceptance thresholds are again $0.99$ and $0.999$ for direct and iterative training, respectively.

Both training methods use Trotter-based initialization and, when compatible with the candidate depth, warm starts from the best parameters at the preceding depth or the selected parameters for $s-1$. Additional layers are initialized with zero angles, while parameters transferred from $s-1$ are rescaled by $s/(s-1)$. One additional initialization is obtained by adding independent Gaussian perturbations with standard deviation $0.02$ to an available warm start, or to the Trotter initialization if no warm start is available.

\subsection{The dissipative quantum \texorpdfstring{$XY$}{XY} model}

We first apply the adjoint channel and adaptive trajectory-compression strategy to simulate a physically motivated open-system transport problem: the melting of a domain-wall state in a dissipative one-dimensional $XY$ chain. The Hamiltonian is given by
\begin{equation}\label{eq:dissi}
    H = \sum_{i=0}^{n-2} \left( X_i X_{i+1} + Y_i Y_{i+1} \right),
\end{equation}
which is equivalent to nearest-neighbor hopping of hard-core particles. We include local dephasing by defining the dissipation as
    \begin{equation}\label{eq:dephase}
    P_k =Z_{k-1}, \qquad \gamma_k = \gamma,  \qquad k=1,2,\ldots,n .
\end{equation}
Thus, the coherent $XY$ Hamiltonian tends to spread local excitations along the chain, whereas the dephasing environment suppresses phase coherence without changing the total excitation number. This model therefore captures the competition between coherent quantum transport and environment-induced classicalization.

Unless otherwise specified, we use $n=10$, $\gamma=0.06$, total evolution time $T=10$, and time step $\delta t=0.1$. The initial state is the half-filled domain-wall product state
\begin{equation}
    \rho(0) = |\psi_0\rangle\langle\psi_0|, \qquad  |\psi_0\rangle = |1\rangle^{\otimes 5}|0\rangle^{\otimes 5}.
\end{equation}
We monitor the domain-wall imbalance
\begin{equation}
    I(t) = \operatorname{Tr}[\hat{I}\rho(t)], \qquad  \hat{I}  = \frac{1}{n}  \left( \sum_{i=0}^{n/2-1} Z_i - \sum_{i=n/2}^{n-1} Z_i \right).
\end{equation}
For the chosen initial state, $I(0)=-1$. As the domain wall melts, particles initially localized on the left half of the chain spread toward the right half, and the imbalance relaxes toward zero. Coherent hopping produces oscillatory transport, while local dephasing damps these oscillations and drives the dynamics toward a more classical relaxation behavior. The imbalance therefore directly probes the nonequilibrium density relaxation of the dissipative chain.

We first test the correctness of the proposed adjoint channel. The deterministic channel result, denoted by ``Kraus,'' is obtained by summing over all possible trajectories at every time step. We also estimate the same channel dynamics by sampling $M$ stochastic trajectories, with $M\in\{32,64,128,256,512\}$. For each $M$, the complete sampling procedure is repeated $100$ times to evaluate the sampling stability.

\begin{figure}[ht]
    \centering
    \includegraphics[width=\columnwidth]{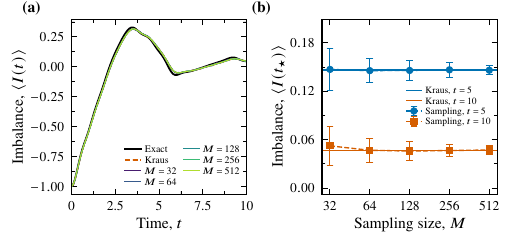}
    \caption{
        \textbf{Simulation results on dissipative domain-wall melting and sampling stability.}
        (a) Domain-wall imbalance obtained from the exact Lindblad equation, the deterministic Kraus channel, and the stochastic trajectory averages for $M=32,64,128,256$, and $512$. Each sampling curve is averaged over $100$ independent repetitions.
        (b) Sampling means at $t=5$ and $t=10$ as functions of $M$. Error bars denote one standard deviation over the $100$ repetitions.
    }
    \label{fig:samplingimbalance}
\end{figure}

As shown in Fig.~\ref{fig:samplingimbalance}(a), the deterministic Kraus evolution follows the exact Lindblad dynamics over the full time window. The sampled trajectories reproduce the same relaxation and oscillation pattern. Figure~\ref{fig:samplingimbalance}(b) further shows the sampling distributions at $t=5$ and $t=10$, where the error bars shrink monotonically with increasing $M$. This confirms that the stochastic adjoint-channel trajectories provide a controlled sampling approximation to the dissipative dynamics.

\begin{figure}[ht]
    \centering
    \includegraphics[width=\columnwidth]{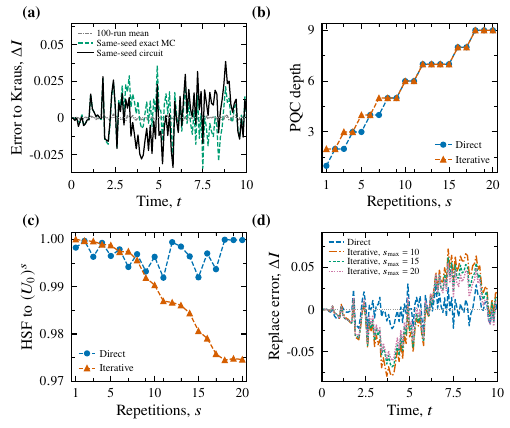}
    \caption{
        \textbf{Adaptive trajectory compression using random product state batches with $B=32$.}
        (a) Signed errors relative to the deterministic Kraus evolution. The panel compares the $100$-run sampling mean, one same-seed exact Monte Carlo trajectory ensemble, and the corresponding same-seed compiled circuit for $M=128$.
        (b) Optimized PQC depth for direct and iterative $B=32$ training as a function of the repetition number $s$.
        (c) Validated HSF of the optimized PQC relative to the exact power $(U_0)^s$.
        (d) Signed replacement errors for the direct library and the iterative libraries with $s_{\text{max}}=10,15$, and $20$. All replacement simulations use $M=128$ trajectories.
    }
    \label{fig:xyb32}
\end{figure}

We then study the gate-level realization of the sampled trajectories and their compression. For an approximate evolution $I_{\mathrm{app}}(t)$, we define the signed imbalance error relative to the deterministic channel as
\begin{equation}\label{eq:xy_imbalance_error}
    \Delta I(t)  =  I_{\mathrm{Kraus}}(t) - I_{\mathrm{app}}(t).
\end{equation}
We perform the following simulations with $M=128$. In Fig.~\ref{fig:xyb32}(a), we first plot the signed error of the mean value over the above $100$ independent sampling repetitions, which remains close to zero throughout the dynamics. We then show another independently sampled trajectory ensemble, which fluctuates around the mean curve with a maximum error of about $0.03$. We also evaluate the same sampled trajectory ensemble after transforming each trajectory into a quantum circuit, where a layer-merging procedure is introduced to simplify the compiled circuit. The circuit-level results closely follow the corresponding sampled curve, indicating that the sampled adjoint-channel dynamics can be implemented at the gate level. The circuit construction and layer-merging procedure are described in Appendix~\ref{sec:xycircuit}.

Next, we compress consecutive applications of the coherent evolution block $U_0$. For a decomposition $H=\sum_i h_i$, we use a Hamiltonian variational ansatz (HVA)~\cite{hva1, hva2}
\begin{equation}
    U(\bm{\theta})  =  \prod_{\ell=1}^{L}  \prod_i  e^{-i\theta_{\ell i}h_i}.
\end{equation}
With the same circuit construction and simplification procedure, one ansatz layer has the same elementary-gate structure as one original Trotter block. Therefore, the comparison of circuit depth between the original and compressed trajectories is fair.

Figure~\ref{fig:xyb32}(b) shows the selected PQC depth as a function of the repetition number $s$. Both direct and iterative training require only slowly increasing depth, reaching depth $9$ at $s=20$. Thus, a long sequence of repeated $U_0$ blocks can be represented by a substantially shorter variational circuit. As mentioned above, the depth of one ansatz layer is comparable to that of one original Trotter block under our circuit construction. Figure~\ref{fig:xyb32}(c)
    plots the HSF between the optimized ansatz and the exact $(U_0)^s$, and
shows that the direct strategy maintains high fidelity over the full range $s=1,\ldots,20$. The iterative strategy follows the target well for short and intermediate blocks, but its fidelity gradually decreases as the teacher errors accumulate.

After training, the PQC library is inserted into the sampled trajectories. Consecutive $U_0$ operations are partitioned into blocks no larger than $s_{\text{max}}$, and each block is replaced by the corresponding optimized PQC. Figure~\ref{fig:xyb32}(d) shows the resulting imbalance errors. Direct replacement preserves the dissipative relaxation curve most accurately. Iterative replacement improves as $s_{\text{max}}$ increases, but remains more sensitive to accumulated training error. Overall, the compressed trajectories reproduce the physical domain-wall melting dynamics while using shorter coherent-evolution circuits.

\begin{figure}[t]
    \centering
    \includegraphics[width=\columnwidth]
    {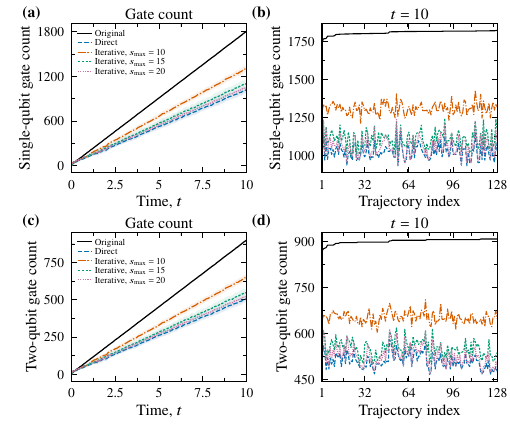}
    \caption{
        \textbf{Circuit-resource reduction from adaptive trajectory compression.}
        (a) Mean single-qubit gate count as a function of time for the uncompressed circuit, direct replacement, and iterative replacement with $s_{\text{max}}=10,15$, and $20$. Shaded regions indicate one standard deviation over $M=128$ trajectories.
        (b) Single-qubit gate counts of individual trajectories at $t=10$, ordered by the corresponding uncompressed count.
        (c) Mean two-qubit gate count as a function of time.
        (d) Two-qubit gate counts of individual trajectories at $t=10$.
    }
    \label{fig:b32gatecount}
\end{figure}

Finally, we quantify the circuit-resource reduction obtained by replacing coherent $U_0$ runs with the trained PQCs. Figures~\ref{fig:b32gatecount}(a) and~\ref{fig:b32gatecount}(c) show the trajectory-averaged single- and two-qubit gate counts. The solid curves denote means over $M=128$ trajectories, and the shaded regions denote one standard deviation. Figures~\ref{fig:b32gatecount}(b) and~\ref{fig:b32gatecount}(d) show the individual trajectory costs at $t=10$, ordered by the corresponding uncompressed gate count.

The direct library gives the largest resource reduction while preserving the imbalance dynamics. At $t=10$, direct replacement lowers both the single-qubit and two-qubit gate counts by about $43\%$ on average. Averaged over all recorded times and trajectories, the reductions remain of the same order, about $42\%$ for both gate types. Iterative replacement also reduces the circuit depth, and its savings increase with $s_{\text{max}}$, but the direct strategy provides the best balance between accuracy and compression in this test.

The simulation with full-space optimization is shown in Appendix \ref{sec:xyfull}, which gives the same selected PQC depths as the $B=32$ calculation and produces the same gate-resource curves. These simulations show that the proposed method is not only a consistency check of the channel construction, but also provides an efficient way to study a physically meaningful open-system transport problem, namely the relaxation of a domain-wall density profile under the competition between coherent $XY$ hopping and local dephasing.

\subsection{The interacting disordered $XXZ$ model}

The above benchmark has a special structure: the $XY$ model Hamiltonian maps to free fermions, and local $Z$ jumps preserve Gaussian evolutions along individual trajectories. This structure can lead to compressions without variational training. For instance, the algorithmic compression in Ref. \cite{kokcu2023} produces circuits with depth independent of evolution time for the free-fermion models. As mentioned above, our simulation framework is not restricted to specific circuit compression strategies. While the above simulation demonstrated the reliability of our method, this structure-aware compilation strategy can also be applied to our method for more efficient trajectory compression. The benchmark with this strategy is shown in Appendix \ref{sec:sa-xy}, which preserves the accuracy while achieving more quantum gate reductions.

To demonstrate our method beyond the free-fermion case, we consider the one-dimensional interacting XXZ model with on-site disorder. The Hamiltonian is
\begin{equation}
    H =\sum_{i=0}^{n-2} \left[  J (X_iX_{i+1} + Y_iY_{i+1}) +J_z Z_iZ_{i+1} \right] + \sum_{i=0}^{n-1} h_iZ_i.
\end{equation}
We take $J=J_z=1$ and independently sample the coefficients $h_i$, uniformly from $[-1,1]$. The $ZZ$ term introduces fermionic density interactions, so the individual trajectories are no longer confined to Gaussian dynamics. We retain the local dephasing convention of Eq. \eqref{eq:dephase}. We set $n=8$, the initial state $|11110000\rangle$, and the other parameters are set the same as in the above simulation. Quantum circuit construction of this model is provided in Appendix \ref{sec:circxxz}.

Figure \ref{fig:samplingimbalancexxz} compares the exact Lindblad evolution, the deterministic adjoint channel, and trajectory sampling. The sampled means follow the deterministic channel, while their spread decreases as the number of trajectories grows. 

\begin{figure}[ht]
    \centering
    \includegraphics[width=\columnwidth]{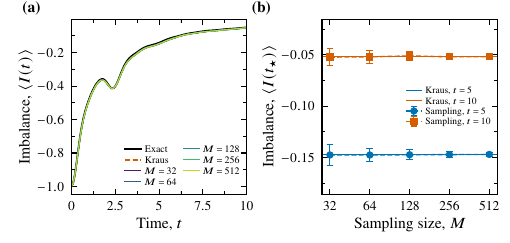}
    \caption{
        \textbf{Domain-wall relaxation in the disordered $XXZ$ chain.}
        (a) Imbalance from the exact Lindblad equation, the deterministic Kraus channel, and sampled trajectories with different $M$. Each sampling curve is averaged over 100 independent repetitions.
        (b) Sampling means at $t=5$ and $t=10$ with one-standard-deviation error bars over the repetitions. 
    }
    \label{fig:samplingimbalancexxz}
\end{figure}

\begin{figure}[ht]
    \centering
    \includegraphics[width=\columnwidth]{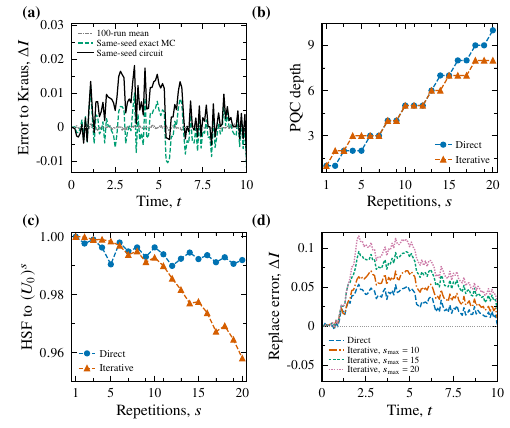}
    \caption{
        \textbf{Adaptive trajectory compression for the disordered $XXZ$ model with $B=32$.}
        (a) Signed imbalance errors relative to the channel for the 100-run sampling mean, one exact Monte Carlo ensemble, and its uncompressed Trotter circuit realization.
        (b) Selected PQC depth, measured in symmetric slices. 
        (c) Validated HSF of the optimized PQC relative to the exact power $(U_0)^s$.
        (d) Signed replacement errors for the direct library and the iterative libraries with $s_{\text{max}}=10,15$, and $20$.
    }
    \label{fig:xxzb32}
\end{figure}

For trajectory compression, we follow the ansatz structure in Ref. \cite{ncvqc} and use a HVA consisting of symmetric slices of diagonal, even-bond, and odd-bond rotations. Figure \ref{fig:xxzb32}(a) first separates the sampling and baseline-circuit errors. Panels (b) and (c) then show the selected depths and the evaluated HSF. Direct training requires increasing depth to approximate longer blocks, reaching 10 symmetric slices at $s=20$. Iterative training selects a shallower circuit, but its fidelity relative to the exact $(U_0)^s$ decreases more noticeably.  After insertion, the compiled libraries still reproduce the overall relaxation, as shown in Fig. \ref{fig:xxzb32}(d). Direct replacement is more accurate than iterative replacement in this instance. In contrast to the $XY$ example, increasing the iterative cutoff does not systematically improve the dynamical accuracy: longer recursively compiled blocks accumulate additional approximation error.

\begin{figure}[t]
    \centering
    \includegraphics[width=\columnwidth]{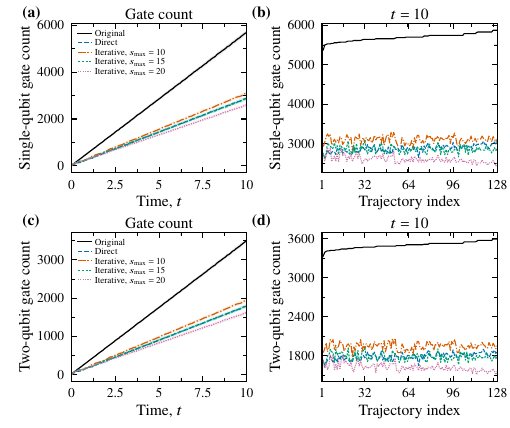}
    \caption{
        \textbf{Execution gate counts for the $XXZ$ product-state-trained libraries.}
        (a,c) Mean single- and two-qubit gate counts for the original trajectories and their direct or iterative replacements. Shading denotes one standard deviation across the same $M=128$ trajectories.
        (b,d) Individual trajectory counts at $t=10$, ordered by the corresponding original count.
    }
    \label{fig:b32gatecountxxz}
\end{figure}

Figure \ref{fig:b32gatecountxxz} shows the corresponding execution gate counts. Direct replacement reduces the mean single- and two-qubit counts by approximately one half at $t=10$. The iterative library can reduce them further, but with the larger dynamical error seen above. Thus the interacting example retains a substantial execution benefit while exhibiting a clearer accuracy–compression tradeoff.

We also performed full-space simulations for this model, and the results are shown in Appendix \ref{sec:xxzfull}. We observe similar selected PQC depth and error trends. To further demonstrate our method under finite shot noise, we ran additional numerical simulations; the results appear in Appendix \ref{sec:finite}. Higher shot budgets bring the selected circuits closer to the noiseless reference, while low-shot stopping can overestimate the achieved fidelity. Additional benchmarks of the system-size dependence of variational compression are presented in Appendix \ref{sec:scalcomp}. Together, these simulations demonstrate the reliability and depth reduction of our proposed method.

\section{Conclusion and Outlook}

In this work, we have studied ancilla-free trajectory simulation of Lindbladian dynamics with adaptive trajectory compression. For open quantum systems with Pauli dissipations, we first derived a compact and stable adjoint channel that directly approximates the dissipative dynamics. Since the resulting channel is mixed-unitary, it can be efficiently implemented on quantum hardware through a trajectory sampling procedure without auxiliary qubits. To further reduce the circuit depth required for implementing each sampled trajectory, we introduced an adaptive variational quantum trajectory compression framework. In this framework, a series of depth-adaptive PQCs are trained to approximate repeated Hamiltonian simulation blocks and are then inserted into the sampled trajectories to replace consecutive no-jump segments. We also discussed different loss-function evaluation and target-construction strategies for training the compressed blocks. In particular, the  random product-state loss provides a hardware-friendly training objective that can be evaluated without auxiliary qubits, and more importantly, other compression strategies can also be applied within our framework based on specific Hamiltonian structure. Numerical results on the dissipative quantum $XY$ model and the interacting disordered $XXZ$ chain demonstrate the reliability of the adjoint channel and the effectiveness of the proposed depth-compression method.

Several remarks are in order. First, the adjoint channel enables an ancilla-free quantum simulation of open quantum systems dynamics, and the adaptive trajectory compression framework is further applied to reduce the circuit depth when implementing the sampled trajectories. However, although the numerical results demonstrate the feasibility of constructing the compressed circuit libraries for the models considered, the associated training overhead does not necessarily lead to an overall reduction in computational resources. We would like to point out that circuit depth reduction in this case is still meaningful, as current NISQ devices are mainly affected by noise. Further work will try to optimize the training procedure in more real cases with circuit approximation and hardware noise taking into account.

Second, the adjoint channel derived in this work has a local approximation error scaling as $\mathcal{O}(\delta t^2)$, leading to a global error scaling as $\mathcal{O}(T\delta t)$ for a simulation time $T=m\delta t$. Quantum channels with higher-order accuracy have been widely studied~\cite{higherorder1, higherorder2, higherorder3}, and in principle they can reduce the algorithmic discretization error. However, higher-order constructions typically generate Kraus operators involving products of $U_0$ and $P_k$, which makes the sampled trajectories structurally more complicated. This is different from the adjoint channel used in this work, where each trajectory is generated only by the simple elementary blocks $U_0$ and $P_k$. In the weak-dissipation regime, $U_0$ appears with high probability and therefore forms long consecutive no-jump segments. This compact trajectory structure is precisely what allows the adaptive variational compression framework to train PQCs only for $(U_0)^s$, rather than requiring independent training for different trajectory-dependent products.

From the perspective of NISQ implementation, the tradeoff between algorithmic accuracy and hardware noise is also essential. Although higher-order channels can reduce the discretization error for a fixed time step, they may require deeper circuits, more complicated trajectory structures, and larger sampling overhead. On current quantum processors, increasing the formal order of the algorithm does not necessarily improve the final simulation accuracy if the additional circuit depth introduces stronger hardware noise. Therefore, for near-term devices, a lower-order but compact, ancilla-free, and depth-reduced algorithm can be more practical than a formally higher-order construction with substantially larger resource requirements. Understanding and optimizing this balance between algorithmic error and hardware-induced error is an important direction for future work.

Overall, our results suggest that combining physically structured adjoint-channel representations with variational trajectory compression provides a practical route toward open-system quantum simulation on near-term quantum hardware. This framework may be further extended to more general dissipative models, higher-order channel constructions, and application-oriented simulations of realistic many-body open quantum systems.

\section*{Acknowledgements}

This work has been supported by the National Key Research and Development Program of China (Grant No. 2023YFB4502500).

\section*{Code and data availability}

The code and data that support the findings of this study are publicly available at \url{https://github.com/WisdomLiuLiu/Paper-supplementary-data/tree/main/avqtc}.

\section*{competing interests}

The authors declare no competing interests.

\section*{Author contributions}

H.-Y. L., C. X., and C. W. proposed the algorithm. H.-Y. L., Y.-J. W., and X.-N. Z. performed the numerical simulations. Y.-C. W. and Z.-Y. C. analyzed the results. H.-Y. L. and Z.-Y. C. wrote the manuscript. G.-P. G. supervised the project. All authors discussed the results and reviewed the manuscript.

\appendix

\section{Hilbert--Schmidt Product Evaluation}\label{sec:hst}

The HSF of two unitary operators $U$ and $V$ is defined as
\begin{equation}
    F_{\text{HS}}(U,V) =\left| \frac{1}{d} \operatorname{Tr}[U^\dagger V] \right|^2 ,
\end{equation}
with $d=2^n$ the Hilbert space dimension. In general, $F_{\text{HS}}(U,V)=1$ implies that $U$ and $V$ are equal up to a global phase, which is irrelevant for expectation values of the form $\operatorname{Tr}[U\rho U^\dagger O]$. Then it can be used to construct the loss function in Eq.~\eqref{eq:losshs}.

This quantity can be evaluated using $2n$ qubits. An example of the corresponding circuit for $n=2$ is shown in Fig.~\ref{fig:hstest}. The circuit consists of two registers, denoted by $a$ and $b$. We now analyze the procedure step by step.

\begin{figure}[ht]
    \centering
    \begin{tikzpicture}
    \node[scale=1]{
    \begin{tikzcd}
    \lstick{$a_0$} & \gate{H} & \ctrl{2} & & \gate[2]{V} & & \ctrl{2} & \gate{H} & \meter{} \\
    \lstick{$a_1$} & \gate{H} &           & \ctrl{2} &      & \ctrl{2} &           & \gate{H} & \meter{} \\
    \lstick{$b_0$} &           & \targ{}  &           & \gate[2]{U^*} &           & \targ{}  &           & \meter{} \\
    \lstick{$b_1$} &           &           & \targ{}  &      & \targ{}  &           &           & \meter{}
    \end{tikzcd}
    };
    \end{tikzpicture}
    \caption{
        \textbf{Quantum circuit for evaluating the HSF.}
    }
    \label{fig:hstest}
\end{figure}

First, we prepare a maximally entangled state between the two registers,
\begin{equation}
    |\Psi\rangle  =  \frac{1}{\sqrt{2^n}}  \sum_i  |i\rangle_a \otimes |i\rangle_b .
\end{equation}
This is achieved by applying a Hadamard gate to each qubit in register $a$, followed by CNOT gates between each pair of qubits $(a_i,b_i)$.

Next, we apply $V$ to register $a$ and $U^*$ to register $b$, resulting in the state
\begin{equation}
    |\Phi\rangle = \frac{1}{\sqrt{2^n}}  \sum_i  V|i\rangle_a \otimes U^*|i\rangle_b .
\end{equation}
Here $U^*$ denotes the complex conjugate of $U$ in the computational basis. For common gate sets, such as single-qubit rotations $R_x(\theta)$ and fixed two-qubit gates such as CNOT or CZ, the operator $U^*$ can be obtained by complex conjugating each gate. In particular, $R_x(\theta)^* = R_x(-\theta)$, while CNOT and CZ are real and remain unchanged.

Finally, the overlap between $|\Phi\rangle$ and $|\Psi\rangle$ is given by
\begin{equation}
    \langle \Psi | \Phi \rangle
    =  \frac{1}{2^n}  \sum_{i,j}  \langle i | V | j \rangle  \langle i | U^* | j \rangle 
    = \frac{1}{2^n} \operatorname{Tr}[U^\dagger V].
\end{equation}

To estimate this overlap, we reverse the state preparation of $|\Psi\rangle$ and measure all qubits. The probability of obtaining $|0\rangle_a^{\otimes n}|0\rangle_b^{\otimes n}$ is
\begin{equation}
    p_0  =  |\langle \Psi | \Phi \rangle|^2 =   F_{\text{HS}}(U,V) .
\end{equation}
Therefore, the HSF can be directly estimated from the probability of the all-zero measurement outcome.

Although this procedure requires auxiliary qubits, it is different from the auxiliary-qubit requirements in dilation-based constructions, LCU-based state preparation, or Hadamard-test-based overlap estimation. In those approaches, two-qubit gates are usually required between the auxiliary qubits and every qubit in the system register, which can lead to relatively deep circuits or require multiple SWAP gates under limited qubit connectivity. In contrast, the HST circuit exemplified in Fig.~\ref{fig:hstest} only requires $2n$ qubits arranged in a $2\times n$ lattice structure, with entangling gates applied locally between the two registers. Therefore, although auxiliary qubits are still needed, the circuit-depth overhead remains moderate.

\section{Quantum circuit implementation details}\label{sec:simcirc}

In this part, we provide details on the circuit construction for implementation of the coherent Hamiltonian evolution operators. Circuit merging strategies are also applied for depth reduction. To fairly compare the depth between the original evolution operator and the ansatz, we use a HVA in these simulations, which shares the same circuit structure between the ansatz and the target Hamiltonian evolution blocks.

\subsection{The dissipative quantum \texorpdfstring{$XY$}{XY} model}
\label{sec:xycircuit}

Here we describe the circuit construction details for simulating the dissipative quantum $XY$ model. First, the Hamiltonian is written as a sum of nearest-neighbor coupling terms,
\begin{equation}\label{eq:hxyappend}
    H = \sum_{i=0}^{n-2} h_i, \qquad h_i = X_iX_{i+1}  + Y_iY_{i+1}.
\end{equation}
For a single term $h_i$, we denote its time-evolution operator as
\begin{equation}
    G_i(t) = e^{-ith_i}.
\end{equation}
Then the Hamiltonian simulation operator $U_0$ in the adjoint channel can be realized via a first-order Trotter decomposition as
\begin{equation}
    U_0 = e^{-iH\Delta t} \approx  \prod_i G_i(\Delta t).
\end{equation}
Since coupling terms acting on disjoint pairs commute, we split $U_0$ into odd and even layers,
\begin{equation}
    U_{\mathrm{o}}(\Delta t)  = \prod_{i=1,3,\cdots} G_i(\Delta t),
    \quad
    U_{\mathrm{e}}(\Delta t) =  \prod_{i=0,2,\cdots} G_i(\Delta t).
\end{equation}
Thus, one Trotter step can be implemented as
\begin{equation}
    U_0 \approx  U_{\mathrm{o}}(\Delta t)U_{\mathrm{e}}(\Delta t),
\end{equation}
or equivalently, up to the same first-order Trotter error, as
\begin{equation}
    U_0 \approx U_{\mathrm{e}}(\Delta t)U_{\mathrm{o}}(\Delta t).
\end{equation}

Inspired by Ref.~\cite{xxyycnot}, each $G_i(\Delta t)$ can be realized by the quantum circuit shown in Fig.~\ref{fig:xyrot}, which uses two CNOT gates.

\begin{figure}[ht]
    \centering
    \begin{tikzpicture}
        \node[scale=0.9] {
        \begin{tikzcd}
    \lstick{$q_i$}
        & \gate{R_x(\pi/2)}
        & \ctrl{1}
        & \gate{R_x(2\Delta t)}
        & \ctrl{1}
        & \gate{R_x(-\pi/2)}
        & \qw \\
    \lstick{$q_{i+1}$}
        & \qw
        & \targ{}
        & \gate{R_y(2\Delta t)}
        & \targ{}
        & \qw
        & \qw
        \end{tikzcd}
        };
    \end{tikzpicture}
    \caption{
        \textbf{Quantum circuit for implementing $G_i(\Delta t)$.}
        It requires 2 CNOT gates and four single-qubit rotation gates.
    }
    \label{fig:xyrot}
\end{figure}

For the gate-level implementation, we approximate $U_0$ using a Trotter decomposition and apply the resulting circuit sequentially.
A direct implementation of $(U_0)^s$ contains $2s$ odd-even layers. However, in the circuit realization of $G_i$ shown in Fig.~\ref{fig:xyrot}, consecutive gates on the same bond can be merged according to
\begin{equation}
    G_i(\Delta t)G_i(\Delta t) =  G_i(2\Delta t),
\end{equation}
which introduces no additional Trotter error for that bond. On the other hand, since both $U_{\mathrm{o}}(\Delta t)U_{\mathrm{e}}(\Delta t)$ and $U_{\mathrm{e}}(\Delta t)U_{\mathrm{o}}(\Delta t)$ are first-order approximations to $e^{-iH\Delta t}$, we can alternate these two orderings to expose neighboring layers acting on the same bonds. This leads to the following layer-merging form:
\begin{equation}\label{eq:u0evenodd}
\begin{aligned}
  &  U_0U_0U_0\cdots \\
    \approx &[ U_{\mathrm{o}}(\Delta t)U_{\mathrm{e}}(\Delta t)][U_{\mathrm{e}}(\Delta t)U_{\mathrm{o}}(\Delta t)][U_{\mathrm{o}}(\Delta t)U_{\mathrm{e}}(\Delta t)] \cdots \\
    =&U_{\mathrm{o}}(\Delta t) U_{\mathrm{e}}(2\Delta t) U_{\mathrm{o}}(2\Delta t) U_{\mathrm{e}}(2\Delta t) \cdots.
\end{aligned}
\end{equation}
Therefore, the circuit depth for realizing $(U_0)^s$ can be reduced from $2s$ layers to $s+1$ layers. The resulting circuit structure is shown in Fig.~\ref{fig:xyevolve}.

\begin{figure}[ht]
    \centering
    \begin{tikzpicture}
        \node[scale=0.9] {
        \begin{tikzcd}
    \lstick{$q_0$}
        & \qw
        & \gate[2]{G_{0}(2\Delta t)}
        & \qw
        & \gate[2]{G_{0}(\Delta t)}
        & \qw \\
    \lstick{$q_1$}
        & \gate[2]{G_{1}(\Delta t)}
        & \qw
        & \gate[2]{G_{1}(2\Delta t)}
        & \qw
        & \qw \\
    \lstick{$q_2$}
        & \qw
        & \gate[2]{G_{2}(2\Delta t)}
        & \qw
        & \gate[2]{G_{2}(\Delta t)}
        & \qw \\
    \lstick{$q_3$}
        & \gate[2]{G_{3}(\Delta t)}
        & \qw
        & \gate[2]{G_{3}(2\Delta t)}
        & \qw
        & \qw \\
    \lstick{$\vdots$}
        & \vdots
        & \vdots
        & \vdots
        & \vdots
        & \qw
        \end{tikzcd}
        };
    \end{tikzpicture}
    \caption{
        \textbf{Quantum circuit for realizing repeated rotations of the $XY$ Hamiltonian.}
        Permuting the even and odd layers allows neighboring layers on the same bonds to be merged, reducing the circuit depth from $2s$ to $s+1$.
    }
    \label{fig:xyevolve}
\end{figure}

In the training stage, we use the Hamiltonian variational ansatz. For the model Hamiltonian in Eq.~\eqref{eq:hxyappend}, the ansatz has the form
\begin{equation}
    U(\bm{\theta})
    =
    \prod_{\ell=1}^{L}
    \prod_i
    G_i(\theta_{\ell i}),
\end{equation}
where $L$ is the ansatz depth and $\theta_{\ell i}$ is the variational parameter associated with the coupling term $h_i$ in the $\ell$-th layer. Its quantum circuit realization has the same elementary structure as the circuit used for realizing $U_0$ sequences, as shown in Fig.~\ref{fig:xyevolve}. Therefore, the optimized ansatz depth can be directly compared with the number of layers in the original uncompressed circuit.

\subsection{The dissipative interacting disordered $XXZ$ model}\label{sec:circxxz}

The quantum circuit implementation for the interacting disordered $XXZ$ model is based on the above result. Specifically, the Hamiltonian can be divided into $XY$ terms, which can be implemented by the even-odd bond shown above, and an extra group consisting of commuting $Z$ and $ZZ$ terms. Taking the second-order Trotter approximation, we obtain the symmetric product \cite{ncvqc} as
\begin{equation}
    U_0 \approx U_d(\Delta t/2) U_e(\Delta t/2) U_o(\Delta t) U_e(\Delta t/2) U_d(\Delta t/2),
\end{equation}
in which $U_o$ and $U_e$ are defined above, and $U_d$ is the rotation generated by $Z$ and $ZZ$ terms in the Hamiltonian.

For the HVA, we can replace the coefficients in the structure with independent angles: $n$ angles for $Z_i$ terms, $n-1$ angles for $Z_iZ_j$ terms, and $n-1$ angles for $XY$ terms, resulting in $3n-2$ parameters per slice. Adjacent commuting diagonal half-slices are merged similar to Eq. \eqref{eq:u0evenodd}

\section{Additional simulation results using the full-space objective}
\label{sec:suppfull}

In this section, we present the corresponding results obtained with the full-space Hilbert--Schmidt objective. This calculation serves as a deterministic reference for the finite-batch training used in the main text. Instead of using the random product-state objective, we optimize the PQC by using the full HSF-based loss function in Eq.~\eqref{eq:losshs}. For direct training, the target is $U_{\mathrm{tar}}=U_0^s$. For iterative training, the teacher target is constructed as $U(\bm{\theta}_{s-1})U_0$, as in the finite-batch calculation. The optimization procedure follows Sec.~IV A. For both training strategies, we report the HSF relative to the exact target $(U_0)^s$.

\subsection{the dissipative quantum $XY$ model}\label{sec:xyfull}

\begin{figure}[ht]
    \centering
    \includegraphics[width=\columnwidth]{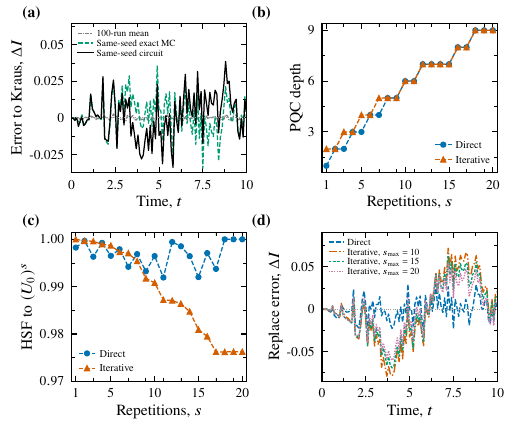}
    \caption{
        \textbf{Adaptive trajectory compression using the full-space objective.}
        (a) Signed errors relative to the deterministic Kraus evolution. The panel compares the $100$-run sampling mean, one same-seed exact Monte Carlo trajectory ensemble, and the corresponding same-seed compiled circuit for $M=128$.
        (b) Optimized PQC depth as a function of the repetition number $s$.
        (c) Validated HSF of the optimized PQC relative to the exact power $(U_0)^s$.
        (d) Signed replacement errors for the direct library and the iterative libraries with $s_{\text{max}}=10,15$, and $20$.
    }
    \label{fig:xyfull}
\end{figure}

We first provide the full-space simulation results for the dissipative quantum $XY$ model. Figure~\ref{fig:xyfull} shows the full-space training and replacement results. Panel (a) repeats the same error-source diagnostic used in the main text. The $100$-run sampling mean remains close to the deterministic Kraus curve, whereas a single same-seed $M=128$ trajectory ensemble already produces a larger finite-sampling fluctuation, with a maximum imbalance error of order $0.03$. The same-seed compiled circuit follows this trajectory-level fluctuation pattern. This confirms that the visible deviations in the gate-level trajectory simulation are mainly associated with finite trajectory sampling, rather than with the PQC training procedure.

The full-space optimization selects essentially the same compact PQC depths as the finite-batch optimization. As shown in Fig.~\ref{fig:xyfull}(b), both direct and iterative training require only slowly increasing depth as the repetition number $s$ grows, reaching depth $9$ at $s=20$. Thus, the compression pattern is not an artifact of the stochastic $B=32$ loss evaluation.

Figure~\ref{fig:xyfull}(c) shows the validation fidelity relative to $(U_0)^s$. The direct full-space training remains highly accurate throughout the tested range and preserves the coherent block well. The iterative training is optimized to match its teacher target at each step, but its fidelity to the exact power decreases gradually with $s$, reflecting the accumulation of teacher errors. This behavior is consistent with the finite-batch results in the main text.

After inserting the full-space trained PQC library into the sampled trajectories, the direct replacement again gives the most stable reconstruction of the dissipative imbalance dynamics, as shown in Fig.~\ref{fig:xyfull}(d). The iterative replacements improve as $s_{\text{max}}$ is increased, but they remain more affected by accumulated training error. Overall, the full-space calculation supports the same conclusion as the $B=32$ calculation: direct training provides the best accuracy for replacing coherent $U_0$ blocks inside the stochastic open-system trajectories.

\begin{figure}[ht]
    \centering
    \includegraphics[width=\columnwidth]{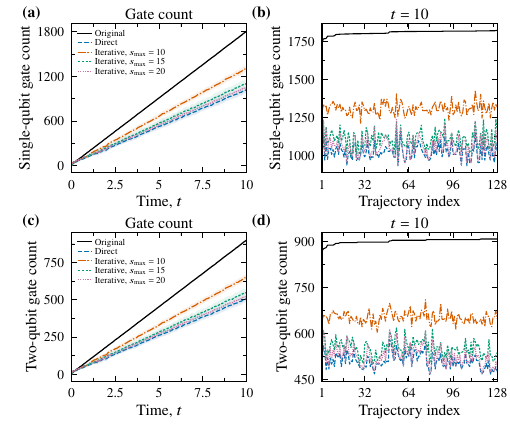}
    \caption{
        \textbf{Circuit-resource reduction for the full-space trained PQC library.}
        (a) Mean single-qubit gate count as a function of time for the uncompressed circuit, direct replacement, and iterative replacement with $s_{\text{max}}=10,15$, and $20$. Shaded regions indicate one standard deviation over $M=128$ trajectories.
        (b) Single-qubit gate counts of individual trajectories at $t=10$, ordered by the corresponding uncompressed count.
        (c) Mean two-qubit gate count as a function of time.
        (d) Two-qubit gate counts of individual trajectories at $t=10$.
    }
    \label{fig:fullgatecount}
\end{figure}

We also examine the corresponding gate-resource reduction. Since the full-space and finite-batch optimizations select the same PQC depths, the resource curves in Fig.~\ref{fig:fullgatecount} show the same overall compression behavior as those in the main text. The direct replacement gives the largest reduction while maintaining the best imbalance accuracy. At $t=10$, it lowers both the single-qubit and two-qubit gate counts by about $43\%$ on average. Averaged over all recorded times and trajectories, the reduction remains about $42\%$ for both gate types. Iterative replacement also reduces the resources, and the saving increases with $s_{\text{max}}$, but this comes with a larger replacement error than the direct strategy.

These full-space results show that the finite-batch $B=32$ optimization used in the main text captures the same physical and circuit-compression trends as the deterministic Hilbert--Schmidt training. The finite-batch method therefore provides a practical route for experimental implementation, while the full-space calculation confirms that the observed compression is not caused by stochastic fluctuations in the sampled loss function.

\subsection{The Interacting disordered $XXZ$ chain} \label{sec:xxzfull}

\begin{figure}[ht]
 \centering
 \includegraphics[width=\columnwidth]{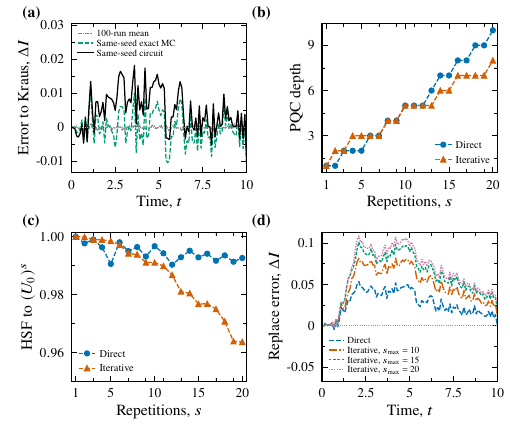}
 \caption{\textbf{Full-space training and replacement for the disordered $XXZ$ chain.}
 Panels show (a) uncompressed sampling and circuit errors, (b) selected slice count, (c) exact HSF relative to $U_0^s$, and (d) replacement error relative to the Kraus channel. The direct cutoff is $s_{\max}=20$, while the iterative cutoffs are 10, 15, and 20. The trajectories and training seed match Fig.~\ref{fig:xxzb32}}
 \label{fig:xxzfull}
\end{figure}

For the disordered $XXZ$ chain, the full-space optimization is performed on the same targets, disorder realization, and trajectory branches as the product-state calculation. Figure~\ref{fig:xxzfull} shows that direct training maintains high HSF, whereas the iterative target introduces accumulated error relative to the exact $(U_0)^s$. At $s=20$, the selected direct and iterative circuits contain ten and eight symmetric slices, respectively. Despite the high teacher fidelity in iterative training, its exact-power fidelity is lower. This confirms that teacher error accumulation is not a consequence of the finite product-state batch alone.

\begin{figure}[ht]
 \centering
 \includegraphics[width=\columnwidth]{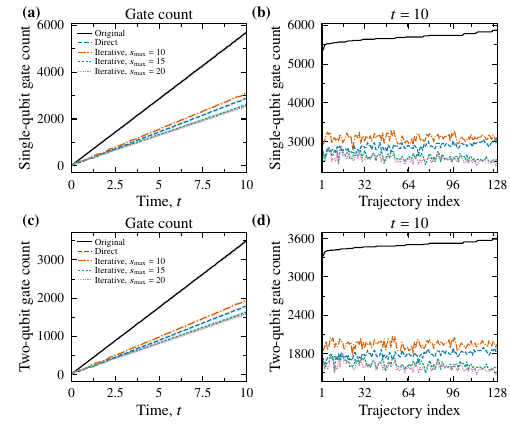}
 \caption{\textbf{Execution gate counts for the full-space-trained $XXZ$ libraries.}
 (a,c) Mean single- and two-qubit gate counts, with one-standard-deviation shading over $M=128$ trajectories. (b,d) Counts at $t=10$, ordered by the original trajectory cost. The library cutoffs and branch samples are those of Fig.~\ref{fig:xxzfull}. Training is excluded.}
 \label{fig:xxzfullgate}
\end{figure}

The gate-count curves in Fig.~\ref{fig:xxzfullgate} likewise follow the product-state results: direct replacement approximately halves the execution cost, while iterative replacement offers further compression at a larger dynamical error. Both replacement strategies have larger imbalance errors than the uncompressed reference. Agreement between the two training objectives supports the qualitative compression trends in this instance.

\section{Training with finite shots}\label{sec:finite}

In the above simulations, parameters are trained with noiseless simulations. To further demonstrate the effectiveness of our algorithm under finite measurement statistics, we provide additional simulations of finite-shot full-space training, in contrast to the random product-state training used in the main text. We consider the $XXZ$ model studied in the main text and independently set the number of shots as $N=256,1024,4096,16384$ in each run. The noiseless simulations are used for comparison.

During optimization, finite-shot estimates of the HST measurement probabilities and their parameter-shift gradients are used to update the circuit parameters. The fidelity is also estimated from measurement outcomes at initialization and after each update. Once the sampled fidelity reaches $0.99$, the current parameters and depth are accepted; otherwise, the adaptive-depth search continues. All optimization and stopping decisions rely on sampled estimates. After the circuit libraries have been fixed, the exact HSF is computed numerically, with an additional independent finite-shot test using $16N$ shots for validation.

\begin{figure*}[ht]
    \centering
    \includegraphics[width=1\linewidth]{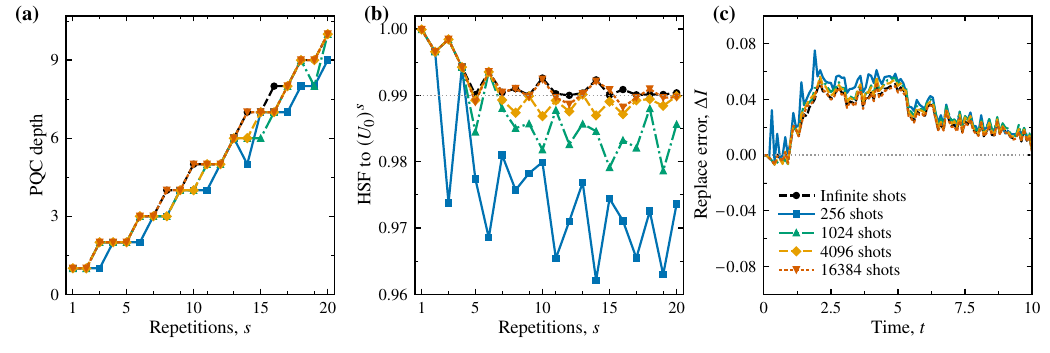}
    \caption{
    \textbf{Finite-shot training and trajectory compression for the interacting disordered $XXZ$ model.}
    (a) Selected PQC depth for a given $s$.
    (b) HSF relative to $(U_0)^s$.
    (c) Relative error to the Kraus model after inserting the library into the trajectories.
    }
    \label{fig:finiteshot}
\end{figure*}

As shown in Fig. \ref{fig:finiteshot}(a) and (b), increasing the shot budget generally brings the selected circuit closer to the noiseless simulation. After training, we insert the five libraries into the same trajectory ensemble. Fig. \ref{fig:finiteshot}(c) shows that higher-shot libraries recover the noiseless replacement behavior more closely. These results show that the simulation can be achieved under finite shot noise, and our future work will consider the effect of real hardware noise.

\section{Additional circuit compression benchmarks}

In this appendix, we present two additional benchmarks for the circuit-compression component of our trajectory simulation framework. We first investigate structure-aware algebraic compression in the dissipative $XY$ model, and then examine the system-size dependence of variational compression in the interacting disordered $XXZ$ model.

\subsection{Structure-aware compression of the dissipative $XY$ model}\label{sec:sa-xy}

As explained in the main text, the $XY$ Hamiltonian and local $Z$ jumps preserve the free-fermionic Gaussian structure along individual trajectories. In addition to variational compression, we therefore consider the structure-aware compression method~\cite{kokcu2023}, which uses fusion, commutation, and turnover identities to rewrite free-fermionic evolution circuits without variational training. We apply this procedure to each consecutive no-jump evolution block in the sampled trajectories. At fixed system size, the compressed depth of an individual coherent block does not grow with its evolution length.

\begin{figure*}[t]
    \centering
    \includegraphics[width=\textwidth]{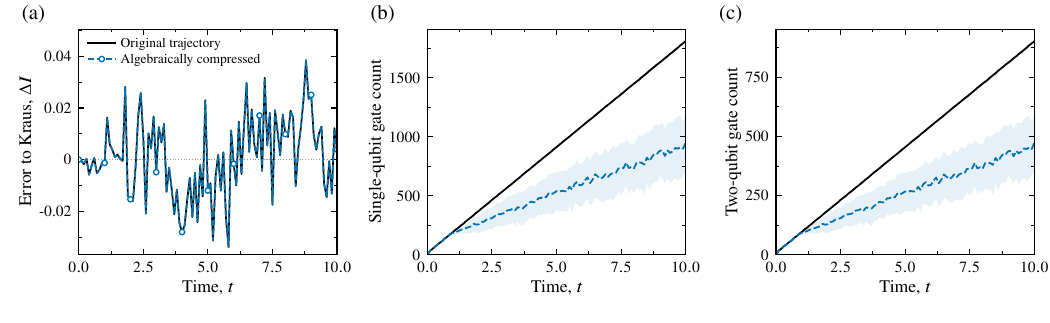}
    \caption{
    \textbf{Algebraic compression for the dissipative $XY$ model.}
    (a) Imbalance errors relative to the deterministic Kraus channel for the original and algebraically compressed trajectory circuits; the curves overlap within numerical precision.
    (b) Mean single-qubit gate count and
    (c) mean two-qubit gate count as functions of simulation time.
    Shaded regions indicate one standard deviation over $M=128$ trajectories. The original and compressed circuits use the same sampled trajectories and simulation parameters.
    }
    \label{fig:xy_algebraic_appendix}
\end{figure*}

Figure~\ref{fig:xy_algebraic_appendix} compares the original and algebraically compressed trajectories for the dissipative $XY$ model with $M=128$. As shown in panel (a), the imbalance errors relative to the deterministic Kraus channel overlap within numerical precision, indicating that the algebraic compression introduces no detectable additional error. The sampling and Trotter errors of the original implementation remain unchanged. Panels (b) and (c) show the trajectory-averaged single- and two-qubit gate counts. At $t=10$, algebraic compression reduces these counts by approximately $48\%$. For comparison, the direct variational replacement in the main text gives reductions of approximately $43\%$ for the same model. Thus, the free-fermionic structure allows an effective alternative to variational compilation within the same trajectory-replacement framework.

\subsection{System-size dependence of variational compression}\label{sec:scalcomp}

\begin{figure}[t]
    \centering
    \includegraphics[width=\columnwidth]{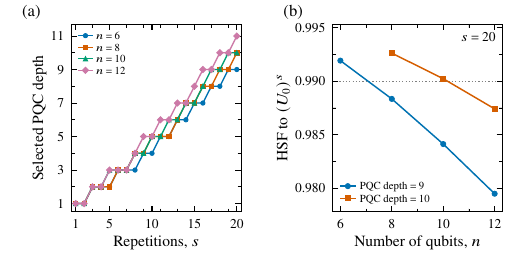}
    \caption{
    \textbf{System-size dependence of variational compression for the interacting disordered $XXZ$ model.}
    (a) Selected PQC depth as a function of the evolution-block length $s$ for $n=6,8,10$, and $12$.
    (b) Independently evaluated HSF relative to $(U_0)^s$ at $s=20$ for fixed PQC depths $9$ and $10$.
    The dotted line indicates an HSF of $0.99$.
    }
    \label{fig:xxz_scaling_appendix}
\end{figure}

To investigate how variational compression depends on system size, we further examine the compression for the interacting disordered $XXZ$ model with $n=6,8,10$, and $12$ qubits. Following the training procedure described in Sec.~\ref{sec:simdetail}, we obtain compressed circuits for evolution-block lengths $s=1,\ldots,20$. Figure~\ref{fig:xxz_scaling_appendix}(a) shows that the selected PQC depth increases gradually with both $s$ and $n$. In particular, at $s=20$, the selected depth increases from $9$ for $n=6$ to $11$ for $n=12$.

To examine the compression accuracy at fixed depth, Fig.~\ref{fig:xxz_scaling_appendix}(b) shows the independently evaluated Hilbert--Schmidt fidelity (HSF) relative to the exact target $(U_0)^s$ at $s=20$, for PQC depths $9$ and $10$. The HSF decreases as the number of qubits increases, while the deeper ansatz generally provides higher fidelity. Although some values fall below $0.99$, the fidelities remain relatively high over the tested system sizes. These results illustrate the tradeoff between ansatz depth and approximation accuracy: larger systems may require additional variational layers to maintain the same target fidelity. The observed gradual increase in selected depth supports the feasibility of compression over the investigated range.

\clearpage
\bibliography{ref}
\end{document}